\documentclass[12pt]{iopart}

\usepackage{graphicx}  
\begin{document}

\title[Prediction and Performance of ITER]{Prediction of Performance and Turbulence in ITER Burning Plasmas via Nonlinear Gyrokinetic Profile Prediction}

\author{N.T. Howard$^1$, P. Rodriguez-Fernandez$^1$, C. Holland$^2$, J. Candy$^3$}
\address{$^1$ MIT Plasma Science and Fusion Center, Cambridge MA 02139, USA}
\address{$^2$ University of California - San Diego, La Jolla, CA 92093, USA}
\address{$^3$ General Atomics, San Diego, CA 92121, USA}
\ead{nthoward@psfc.mit.edu}
\vspace{10pt}
\begin{indented}
\item[]April 2024
\end{indented}

\begin{abstract}
Burning plasma performance, transport, and the effect of hydrogen isotope (H, D, D-T fuel mix) on confinement has been predicted for ITER baseline scenario (IBS) conditions using nonlinear gyrokinetic profile predictions.  Accelerated by surrogate modeling [P. Rodriguez-Fernandez NF 2022], high fidelity, nonlinear gyrokinetic simulations performed with the CGYRO code [J. Candy JCP 2016], were used to predict profiles of $T_i$, $T_e$, and $n_e$ while including the effects of alpha heating, auxiliary power (NBI + ECH), collisional energy exchange, and radiation losses.  Predicted profiles and resulting energy confinement are found to produce fusion power and gain that are approximately consistent with mission goals ($P_{fusion} = 500$MW at $Q=10$) for the baseline scenario and exhibit energy confinement that is within $1\sigma$ of the H-mode energy confinement scaling.  The power of the surrogate modeling technique is demonstrated through the  prediction of alternative ITER scenarios with reduced computational cost.  These scenarios include conditions with maximized fusion gain and an investigation of potential Resonant Magnetic Perturbation (RMP) effects on performance with a minimal number of gyrokinetic profile iterations required (3-6).  These predictions highlight the stiff ITG nature of the core turbulence predicted in the ITER baseline and demonstrate that Q $>$ 17 conditions may be accessible by reducing auxiliary input power while operating in IBS conditions.  Prediction of full kinetic profiles allowed for the projection of hydrogen isotope effects around ITER baseline conditions.  The gyrokinetic fuel ion species was varied from H, D, and 50/50 D-T and kinetic profiles were predicted.  Results indicate that a weak or negligible isotope effect will be observed to arise from core turbulence in ITER baseline scenario conditions.  The resulting energy confinement, turbulence, and density peaking, and the implications for ITER operations will be discussed. 
\end{abstract}

%
%
%
%
%

\section{Introduction}
It is widely accepted that the development of practical fusion energy would provide a source of clean energy that could play a crucial role in mitigating the effects of climate change.  The urgent need for the development of clean energy sources has spurred increased interest in the development of both publicly and privately funded next-generation fusion devices.  The most well-known of these next generation fusion reactors, the ITER tokamak \cite{iter}, is currently under construction in the southern France as part of an international collaboration for the development of fusion.  ITER is believed to be capable of accessing burning plasma conditions, where the self heating from fusion generated alpha particles is equal to or greater than the external heating applied to sustain the fusion reactions.  Confinement and performance in the core of tokamak fusion devices is known to be limited by plasma turbulence that is driven unstable by gradients in the plasma profiles.  As a result, high fidelity modeling capabilities have been developed that allow for the prediction of plasma turbulence in the conditions and geometry typically found in fusion devices.   The most physically comprehensive of these models, known as gyrokinetics, has been validated against experimental measurements on tokamaks worldwide for over two decades and is generally accepted as being an accurate description of the turbulence and transport found in modern tokamaks.  Nonlinear gyrokinetic simulations generally require the use of high performance computing (HPC) and therefore gyrokinetics has typically been applied to study single radial locations for the analysis of dedicated experiments.  However, 3 developments have opened up new possibilities to expand beyond single radial investigations using gyrokinetics: 1. Recent advances in HPC including the widespread usage of hybrid CPU/GPU systems, 2. optimization of gyrokinetic codes on GPU-based HPC platforms and 3. the rapid adoption of machine learning tools.  With these advances, the possibility exists for performing nonlinear gyrokinetic simulations with relatively rapid turnaround with high physics fidelity, which in turn allows for more routine prediction of kinetic profiles and even the optimization of tokamak operation using direct nonlinear gyrokinetic simulation.   In this paper, we describe the use of nonlinear gyrokinetic simulations to predict and optimize the performance of the ITER tokamak, to understand the turbulence expected to dominate ITER baseline conditions, and to probe the potential effect of hydrogen isotope on ITER operation and energy confinement.  The remainder of this paper is organized as follows: Section 2 covers the numerical setup utilized in this work including both the gyrokinetic simulation details and the machine learning framework that enabled more rapid convergence to steady state profiles.  Section 3 covers the kinetic profile predictions and resulting performance for 3 ITER scenarios: the ITER baseline scenario (IBS), a potential enhanced Q scenario, and an investigation of performance in ITER operated with applied resonant magnetic perturbations (RMPs).  Section 4 describes a study of the isotope effect in ITER baseline conditions.  Section 5 briefly describes the conclusions and provides a discussion of the results.

\section{Description of ITER Conditions, Gyrokinetic Simulation Setup, and the Profile Prediction Framework}
\subsection{Description of the ITER conditions}
This paper focuses on the analysis of the ITER baseline scenario (IBS) that has been outlined in previous ITER publications as one of the machine's target operational scenarios \cite{shi07}.   The objective of this condition is to obtain burning plasma conditions with a plasma gain (Q) of 10 while generating 500MW or more power via D-T fusion \cite{iter}.  This will be done by operating the machine with 15MA of plasma current and with a $q_{95}$ of $\sim$3.0.  This scenario has been studied extensively in the literature with a range of experiments and reduced models for the prediction of the core profiles \cite{gri18,cit23,hol23}.  The starting point of this work is JINTRAC modeling that was performed in reference \cite{jintrac} of the ITER baseline scenario.  The output from this modeling was later utilized for modeling of core profiles and performance in References \cite{man20}, \cite{gri18}, and \cite{hol23} utilizing reduced fidelity transport models such as Qualikiz \cite{bou07} and TGLF \cite{sta16}.  In these later works, predictions of toroidal rotation were performed using TGLF-SAT2 and the pedestal pressures were updated to be consistent with predictions from EPED \cite{sny11}.  The results of this TGLF SAT2 work were used at the starting point for our analysis.  We note that published results predicting performance using models such as TGLF and QualiKiz tend to agree reasonably well with a $Q\sim 10$ operational point for the ITER baseline scenario \cite{hol23,cit23}.  

\subsection{Setup of the Gyrokinetic Simulations}
\begin{figure}[h]
\centering
\includegraphics[width=0.6\linewidth]{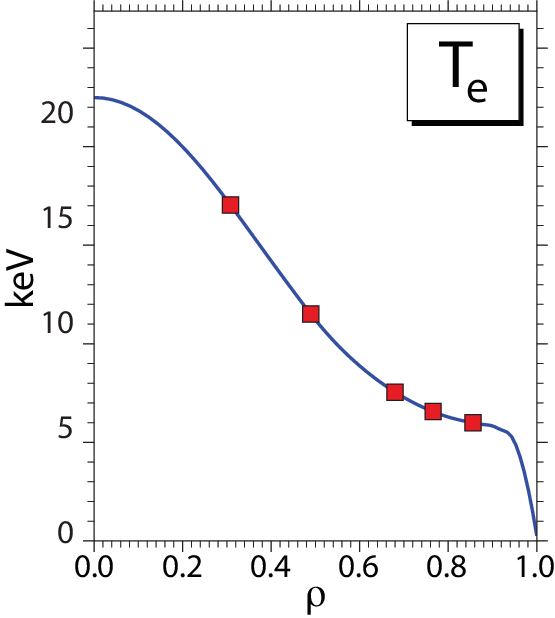}
\caption{\label{locs}(Color Online) An example of an ITER predicted $T_e$ profile plotted against the radial coordinate square root of normalized toroidal flux.  The radial locations of the gyrokinetic evaluations indicated by the red squares, demonstrated a high concentration of points in the outer half of the plasma. }
\end{figure}

All simulations performed in this paper utilized the mature gyrokinetic code CGYRO \cite{can16}.  CGYRO is a local, fully-spectral Eulerian gyrokinetic code optimized for modern computing architectures.   All of the simulations performed for this work were ion-scale (low-k) in nature, capturing turbulence in the range $k_\theta \rho_{s,D} = [ 0 - 1.219]$ and evolved both electrostatic ($\delta \phi$) as well as electromagnetic perturbations ($\delta A_{||}, \delta B_{||}$).  Here $k_\theta$ is the poloidal wavenumber of the turbulence, and $\rho_{s,D}$ is the ion Larmor radius for deuterium evaluated with the ion sound speed.  Although typical simulation domains [$L_x$,$L_y$] varied  slightly based on the radial location simulated, typical simulation box sizes of approximately [120 x 120$\rho_{s,D}$] were utilized and were represented using approximately 512 radial modes ($n_x$), 24 toroidal modes ($n_n$), 24 theta points ($n_\theta$), 24 pitch angles $(n_\xi)$ and 8-12 energies ($n_{energy}$).  All simulations utilized Miller Extended Harmonic (MXH) geometry \cite{arb21}, the Sugama collision operator \cite{sug09}, and retained rotation effects such as ExB shearing.  Unless otherwise specified in the text all simulations performed evolved 5 gyrokinetic species:  Deuterium, Tritium (with a 50/50 ratio), a lumped impurity (Z=6, A=12 ; representing He-ash, Be and Ne contributions), W (1.5e-5 $\times n_e$ ,partially ionized, Z=50, A=184), and electrons.  The combination of the simulation setup used makes these simulations capable of accurately capturing low-k ($k_\theta < 1.0$) turbulence that is believed to be a primary driver of heat and particles losses in many tokamak conditions.  This includes turbulence due to modes such as Ion Temperature Gradient (ITG) modes, Trapped Electron Modes (TEM), and Microtearing modes (MTM).  Simulations were restarted with new profiles/gradients starting from the saturated state of a nonlinear CGYRO simulation run with the initial (TGLF SAT2 \cite{sta21}) profiles.  Each new set of conditions was typically run for $\sim$ 450 $a/c_s$ and averaged heat and particle fluxes were obtained from the last $\sim$ 300 $a/c_s$ of the simulation.  This choice in averaging time represents a balance of heat and particle flux accuracy and computing time usage.  In practice, the simulated heat fluxes were relatively steady in time resulting in typical $1\sigma$ uncertainties of $\sim 10\%$.  All of the simulations presented in this paper were performed on the GPU partition of the NERSC Perlmutter supercomputer with each simulation (a single radial location) typically utilizing 12 nodes, with each node comprised of 4, NVIDIA A100 (40GB) GPUs and requiring approximately 3 hours for completion.  A total of 14 iterations of the code were needed for the initial convergence of the ITER baseline profiles which resulted in 70 total nonlinear gyrokinetic evaluations (5 radial locations x 14 iterations).  
\\
\indent
Inclusion of both low and high-k turbulence was not attempted due to the extreme computational cost of such simulations \cite{how16-nf} and physical arguments based on the power balance heat fluxes in the ITER baseline scenario.  As will be shown in the following sections and as is argued in References \cite{how21-pop, hol23, how24}, the so-called 'fingerprints' argument presented by Kotschenreuther and colleagues \cite{kot19} suggests that plasmas dominated by ion heat flux ($Q_i/Q_e > 1.0$) will likely be dominated by low-k turbulence, most likely ITG at the relatively low values of beta found in in IBS.  As the reader can see from Figure \ref{fluxes} the predicted ion to electron heat flux ratio in the ITER baseline exceeds 1.0 at all locations studied.  This is primarily due to the fact that despite significant energy deposition to the electrons (via electron cyclotron heating and fusion alphas) the collisional energy exchange in these conditions is strong and radiation, that represents only an energy sink only for the electrons, is also high, resulting in a $Q_i/Q_e > 1.0$.  This same result has been shown on several reactor relevant plasmas in Reference \cite{hol23} and on SPARC in Reference \cite{prf22-nf}.  It often stated in the community that burning plasmas will exhibit strong electron dominance since alphas primarily slow down on electrons.  However, this is a naive assessment of the situation, as the strong coupling required for most burning plasma regimes and the high radiation loss at temperatures $> 15 keV$ make this statement not generally correct for many burning plasma conditions.
\\
\indent
Simulations were performed at 5 radial locations that span from $r/a = 0.35 - 0.9$. Radial locations of r/a = 0.35, 0.55, 0.75, 0.825, and 0.9 were used throughout this work (As shown for the $T_e$ profile in Figure \ref{locs}) and were chosen based on analysis of a database of temperature and density profiles on Alcator C-Mod.  These radial locations correspond to (square root of normalized toroidal flux) $\rho = 0.308, 0.488 , 0.681, 0.764, 0.858$ . It was shown in this recent work \cite{prf24} that discretization of $n_e$ and $T_e$ profiles with 5 points, with smooth interpolation between points and tighter grouping of points in the edge region ($r/a > 0.75$) was sufficient to reproduce measured stored energy and fusion powers for ITER like conditions to approximately $5\%$.  The lack of a point inside of $r/a = 0.35$, plays a negligible role in the calculation of the fusion power and stored energy.  Despite this region typically exhibiting the highest temperatures and densities, the volume of the flux surfaces goes to zero on axis and therefore the total contribution to the fusion power is dominated by locations outside this region.

\subsection{Description of the Profile Prediction Framework}
\begin{figure}[h]
\centering
\includegraphics[width=0.5\linewidth]{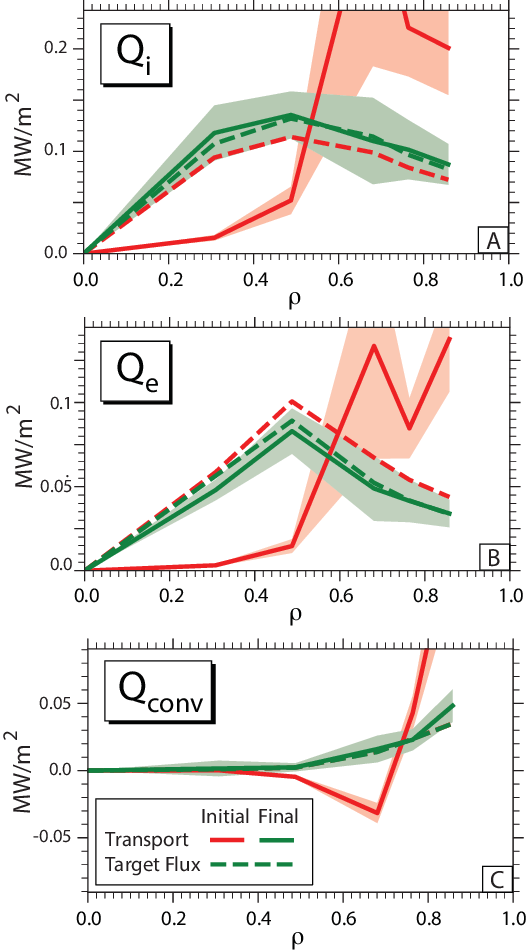}
\caption{\label{fluxes}(Color Online) The $Q_i$ (A), $Q_e$ (B), and $Q_{conv} = 5/2 T_e \Gamma_e$ (C) profiles are plotted for the initial (red) and final (green) conditions obtained during prediction of the ITER baseline condition.  Shaded regions indicate estimated $2\sigma$ uncertainties. }
\end{figure}

All profile predictions presented in this paper utilized the PORTALS \cite{prf24} framework that implements methods for surrogate-accelerated profile prediction.  This framework has been described in detail in Reference \cite{prf22-nf} and a validation of this framework against ITER Similar Shape (ISS) experiments on DIII-D has recently been published in Reference \cite{how24}.  We describe some key features of this framework here for completeness but point the reader to the more comprehensive references above for more details.  Prediction of kinetic $n_e$, $T_e$, and $T_i$ profiles is done by solving the coupled set of energy and particle conversation equations for the discharge of interest.  In this framework, the transport is provided by a sum of turbulent fluxes obtained from nonlinear gyrokinetic simulations (CGYRO) and neoclassical fluxes obtained from running the NEO code \cite{neo}.  The objective of PORTALS is to iterate the density and temperature profiles until the simulated transport matches all target flux values ($Q_i$, $Q_e$, and $\Gamma_e$) simultaneously.  During this work, the temperature and density profiles outside of the outermost simulated radii (r/a = 0.9) are fixed and not evolved during the convergence process. The convergence process initiated through simulation of the initial profiles (obtained from TGLF SAT2 modeling) at the 5 radial locations.  As demonstrated in Figure \ref{fluxes} the initial profiles were in significant disagreement with the target fluxes.  For the next 4 iterations, a simple gradient relaxation method was employed to try to converge to the target fluxes.  The primary purpose of this step is to build up a small database of simulations that can be used to fit a surrogate model.  In this context, a surrogate model is a statistical model that is designed to mimic the properties of the higher fidelity (nonlinear gyrokinetic and neoclassical) models - for example: the relationship of heat and particle fluxes to input parameters such as $a/L_{T_i}$, $a/L_{T_e}$, $a/L_{n}$, etc.  Surrogate models can be rapidly executed and are then used to predict the flux-matched conditions and the resulting profiles.  After the surrogates are used to predict flux matched conditions, high fidelity, nonlinear gyrokinetic simulations were then used to evaluate the predicted profile.  If the gyrokinetic + neoclassical simulated fluxes match the targets, the process stops.  If not, the new high fidelity gyrokinetic simulations are fed back into the database, a new surrogate model is fit to the data, and the process continues until convergence in the high-fidelity models (nonlinear gyrokinetics in our case) is achieved.  Convergence for this work was defined as agreement between the simulated and target fluxes within the $2\sigma$ uncertainties in the simulated values.  Uncertainties in the simulation outputs are the standard deviation of the mean of the time series where the number of total unique samples during the time window is determined using information about the characteristic time of the turbulence fluxes.  An in-depth description of the method is found in Reference \cite{prf22-nf}.  In practice, most radial locations were matched within $1\sigma$ with all locations within $2\sigma$ as shown in Figure \ref{fluxes}.  This technique has been demonstrated to result in convergence approximately 4-6x faster than more standard methods based on Newton solvers \cite{can09,bar10}.  All results shown in this paper are the results of nonlinear gyrokinetic simulations and not from the surrogate themselves. For this work, the geometry was fixed and the surrogates were fit to the known turbulence-relevant quantities $a/L_{T_i}$, $a/L_{T_e}$, $a/L_n$, $T_e/T_i$ and $\nu_{ei}$. At each iteration, PORTALS utilizes the auxiliary heating, ohmic heating and particle source profiles obtained from the original JINTRAC modeling and makes the assumption that these profiles are fixed through the iterations.  Collisional exchange, alpha heating, and radiation (utilizing fits to ADAS\cite{adas} rates) are all self-consistently calculated for each iteration and used as targets to match the transport (neoclassical + nonlinear gyrokinetic) fluxes.

\section{Prediction of Profiles, Turbulence, and Performance for the ITER Baseline Scenario and Potential Paths Towards Optimization}
\begin{figure}[h]
\centering
\includegraphics[width=0.9\linewidth]{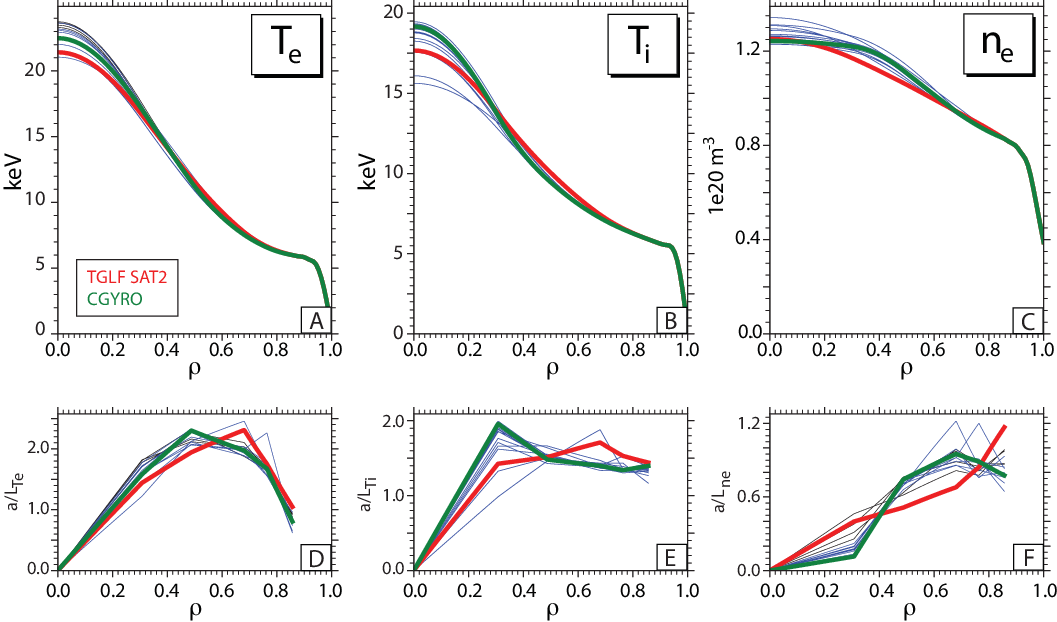}
\caption{\label{itsum}(Color Online) Profiles of $T_e$ (A), $T_i$ (B), and $n_e$ (C), $a/L_{T_e}$ (D), $a/L_{T_i}$ (E), $a/L_{n_e}$ (F) for the convergence of the IBS condition are plotted.  The red curve corresponds to the initial profiles, the green curves correspond to the flux-matched conditions, and the blue lines are variations of the profiles that were tested during convergence.}
\end{figure}

The approach described in Section 2. was applied to predict the profiles and performance of the ITER Baseline Scenario.  Convergence of these profiles required 14 total iterations x 5 radial locations, resulting the need for a total of 70 nonlinear CGYRO runs.  Figure \ref{itsum} provides a summary of this process.  In this Figure we plot the initial (red) and converged (green) $T_e$, $T_i$, and $n_e$ profiles (A-C) as well as their corresponding normalized gradient scale lengths (D-F).  We note that the initial profiles used in this work were taken from the TGLF SAT2 \cite{sta21} modeling described in Reference \cite{hol23}.  The blue curves represent all of the profiles that were tested during the convergence process. 

The nonlinear gyrokinetic simulation predicts that ITER will obtain core electron temperature profiles of approximately 23 keV with ion temperature profiles of approximately 19 keV range on axis with a core density of approximately 1.2e20 $m^{-3}$.  The ion and electron temperatures are well equilibrated from approximately $\rho = 0.4$ and outwards.  Notably, the results of the initial and final profiles are in fairly good agreement.  This implies similarity between the predictions from TGLF SAT2 and CGYRO in these conditions and should provide some additional confidence in TGLF SAT2 based predictions of ITER conditions.  However, it is important to note that this level of agreement should not be considered general.  As shown in previous studies using TGLF and CGYRO profile predictions found significant differences \cite{prf22-nf} in particle transport for the SPARC Primary Reference Discharge.  In Figure \ref{fluxes}, we find that despite the fairly close agreement between initial and final profiles, the initial profiles actually resulted in fluxes that were in poor agreement with the targets.  The fact that the initial and final kinetic profiles were in relatively close agreement points, but the initial fluxes were poorly matched to their target values, is an indication of the stiff nature of the transport in the IBS.

\begin{table}[h]
\caption{\label{tabsum} This table provides a summary of the performance metrics derived from the converged predictions for the ITER baseline scenario and other variation discussed in this paper}
\footnotesize\rm
\begin{tabular*}{\textwidth}{@{}l*{15}{@{\extracolsep{0pt plus12pt}}l}}
\hline
\hline
 & & & ITER Baseline (DT) & & ITER Q Opt.& & ITER RMP& & & & & & &\\
 \hline
 $P_{input}$ & MW & & 53 & & 29 & & 53 & & & & & & &\\
 $P_{fusion}$ & MW & & 498 & & 489 & & 316 & & & & & & &\\
 $Q_{plasma}$ & & & 9.43 & & 16.82 & & 5.98 & & &  & & & &\\
 $\tau_e$ & s & & 2.22 & & 2.62 & & 2.33 & &  & & & & & \\
 $H_{98}$ & & & 0.89 & & 0.93 & & 0.86 & &  & & & & & \\
 $n_e(0.2)/<n_e>$ & & & 1.31 & & 1.32 & & 1.46 & &  & & & & & \\
 $P_{SOL}$ & MW & & 101 & & 76 & & 83 & & & & & & & \\
 $P_{SOL}/P_{LH}$ & & & 1.43 & & 1.08 & & 1.44 & &  & & & & & \\

\hline
\hline
\end{tabular*}
\end{table}

\begin{figure}[h]
\centering
\includegraphics[width=0.8\linewidth]{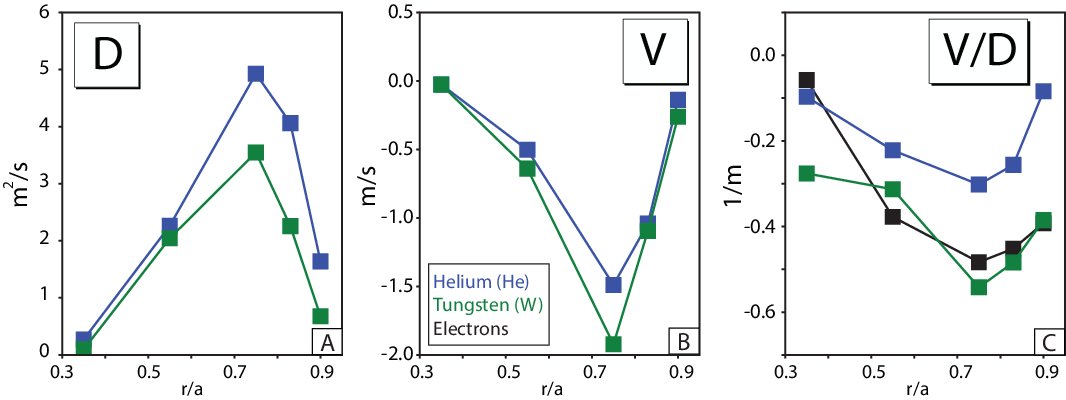}
\caption{\label{imp}(Color Online) The simulated impurity diffusion (A), convection (B), and peaking factor (C) are plotted for trace He and W impurities in the ITER baseline condition.  The peaking factor for the electron profile is also plotted in panel C to indicate that no significant peaking of He or W is expected.}
\end{figure}

Table \ref{tabsum} provides a brief summary of some of the 0-D quantities that were obtained from the converged profiles plotted in Figure \ref{itsum}.  Most notably, the predictions of performance suggest that the ITER baseline will generate 498MW of fusion power with approximately 53MW of total input power ($P_{ECH}+P_{NBI}+P_{OH}$).  This scenario translates to a plasma gain, Q = 9.43 which would put the ITER baseline well within the realm of burning plasma conditions (typically defined as $Q \geq 5$) and very close to its stated goal of Q=10 conditions with 500MW of total fusion power.  Overall, these results are promising for the ITER baseline scenario.  Additionally, it is shown in Table \ref{tabsum} that this condition is operated with an $P_{SOL}$ of 101MW and a $ P_{SOL}/P_{LH}=1.43$ where $P_{LH}$ has been determined from the Martin scaling \cite{mar07} with corrections for isotope mass (m(amu) = 2.5) included.  Therefore these conditions are well above the predicted L to H threshold power.  The density peaking from these profile predictions can also be compared with the scalings calculated by Angioni and colleagues for a database of Alcator C-Mod, ASDEX-Upgrade, and JET H-modes \cite{ang09}.  This plasma condition is operated with an effective collisionality, $\nu_{eff}=0.2 n_{e} R_{geo} T_e^{-2}$ of 0.11 where $n_e$ is in units of $10^{19} m^{-3}$, R is in meters, and $T_e$ is in keV.  The total peaking as defined in Reference \cite{ang09} as $n_e(0.2)/<n_e>$ is found to be 1.31 for the IBS conditions.   We note that this level of peaking is just within the scatter of the density peaking database for the collisonality of interest, but notably at the bottom edge of the database.  In contrast, previous SPARC PRD gyrokinetic predictions had indicated excellent agreement with the density peaking scaling \cite{prf22-nf}.  However, it is important to note that, unlike SPARC, the ITER predictions contain core particle sources arising from both NBI heating and modeled pellet injection.  The derived energy confinement time from the predicted kinetic profiles is 2.22 seconds with a H factor (energy confinement time normalized to the $\tau_{ITER98(y,2)}$ scaling) of 0.89, well within the uncertainty from empirical database ($\sim 15\%$).  Overall, the performance of the ITER baseline appears to be well in line with projections of performance from 0-D empirical modeling and from stated mission goals.
\\
\indent
The successful prediction of kinetic profiles for the ITER baseline motivated an investigation into the nature of turbulent driven impurity transport in these conditions.  Trace impurity species were introduced into the converged simulations at $10^{-6} \times n_e$ to allow for the evaluation of diffusion and convective impurity transport around the converged conditions.  The method used for this work has been described here \cite{how12-nf,  how21-nf}.  This approach was used to evaluate impurity transport for both He and W species as these are known impurities that will be present in ITER and whose transport will play a crucial role in the success of the device.   The derived impurity diffusion (D), convection (V), and peaking factor (V/D) are plotted in Figure \ref{imp}.  The takeaways from this investigation are that the diffusion from these impurities increases going from He to W (with impurity charge), consistent with observations for ITG dominated transport as shown in the DIII-D ITER Similar Shape (ISS) conditions \cite{how24}.  The impurity convection appears to be generally unchanged across the radial locations studied, and there is some trend in the impurity peaking (V/D) with Helium being less peaked (less negative) than tungsten.  However, despite this trend, neither impurity is found to exhibit peaking that would be of a concern in the radial region studied (0.35 - 0.9).  In fact, when compared with the V/D for the electron density profiles ($V/D \sim 1/L_{n_{e}}$), it is seen that the peaking of He and W is comparable or smaller.  We note that neoclassical contributions near axis could potentially lead to non-negligible peaking but such investigations are out of the radial range studied and therefore not presented here.

\begin{figure}[h]
\centering
\includegraphics[width=0.95\linewidth]{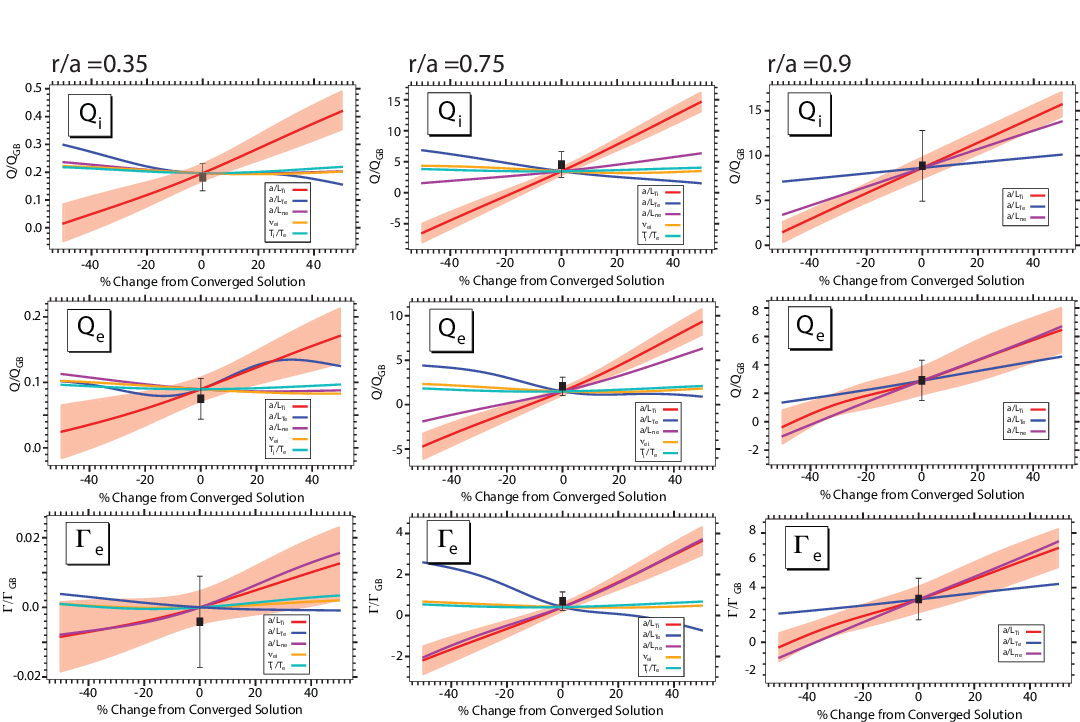}
\caption{\label{sens}(Color Online) The trained surrogates model for the response of the heat and particle fluxes to changes in inputs are shown at 3 radial locations.  Left to right columns are r/a = 0.35, 0.75, and 0.9.  Representative uncertainties are plotted the model for $a/L_{T_i}$ in each panel. }
\end{figure}

We are able to utilize the results of the final surrogate model to investigate the turbulence in these conditions and obtain an understanding of the sensitivity of the heat and particle fluxes to changes in the inputs.   Figure \ref{sens} plots the surrogates (Gaussian Processes) obtained from the converged profiles at 3 of the 5 total radial locations, $r/a = 0.35, 0.75, 0.9$  The surrogates represent a model of how the gyrokinetic simulated heat and particle fluxes are expected to change with changes to the input parameters ($a/L_{T_i}$, $a/L_{T_e}$, $a/L_{n_e}$, $T_e/T_i$, $\nu_{ei}$).  This provides similar information as one would obtain from a single parameter scan performed as part of more traditional gyrokinetic modeling work.  The surrogate uncertainties are plotted only on the $a/L_{T_i}$ curves but are representative of the level of uncertainty found on all parameters scanned.  At r/a = 0.35, $Q_i$, $Q_e$, and $\Gamma_e$ are all most sensitive to changes in $a/L_{T_i}$, the ITG drive term, with generally very weak response of the fluxes to other gradients, particularly in the ion heat flux. Some response of $Q_e$ to changes in $a/L_{T_e}$ is observed and $a/L_{n_e}$ plays a non-negligible role in setting the particle flux at r/a = 0.35.  Note that the surrogates have no knowledge of physics constraints, like the critical gradient for different turbulence types.  As a result, surrogate predicted heat fluxes can extend to significantly negative values (for example in Figure \ref{sens} middle column).  This indicates that high fidelity (CGYRO) model evaluations have likely not been performed in that region of parameter space and the surrogate is extrapolating.  In general the behavior of the surrogates will be most accurate around the converged solution, as that is where the model has the largest amount of data.  The implementation of physics-informed surrogates will be the subject of future work.  At r/a = 0.75 the conclusion is quite similar, with $a/L_{T_i}$ the primary driver of heat and particle fluxes, but with sensitivity to changes in $a/L_{n_e}$ present for the electron heat and particle fluxes.  Results from r/a = 0.9 are plotted in Figure \ref{sens} in the right column.  Since this location was the anchor point for the profile predictions, the absolute values of the profiles ( $n_e$, $T_e$, and $T_i$) are fixed at this location and therefore only three parameters were used to match the local heat fluxes: $a/L_{T_i}$, $a/L_{T_e}$ and $a/L_{n_e}$.  Changes in $a/L_{T_i}$ are again the dominant sensitivity in all of the fluxes, with non-negligible response to changes in $a/L_{T_e}$ and $a/L_{n_e}$ also observed.  This likely is indicative of strong ITG at this radial location with a mix of TEM.  Given the larger minor radius and the larger fraction of trapped particles, an increasing role of TEM would likely be expected at larger major radii.  However, the general conclusion of this modeling is that ITG turbulence plays a dominant role across the radial profile of the ITER baseline conditions.  This ITG dominance can have important implications for reactor operation. As will be demonstrated in the proceeding sections, changes in heating power result in minimal changes to the predicted $T_i$ profile.   From a modeling standpoint, this is a favorable conclusion as it suggests that single scale gyrokinetic simulations are likely sufficient for capturing the anticipated heat and particle transport. 

\subsection{Rapid Optimization and Prediction of ITER Scenarios}
\begin{figure}[h]
\centering
\includegraphics[width=0.9\linewidth]{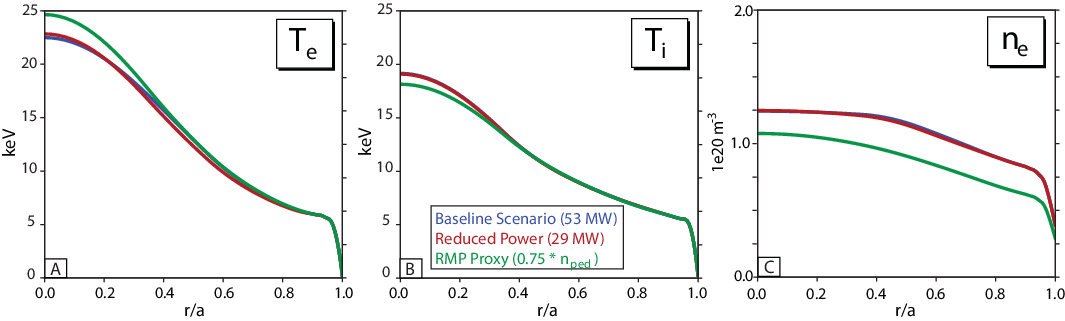}
\caption{\label{itall}(Color Online) Profiles of $T_e$ (A), $T_i$ (B), and $n_e$ (C), as predicted via nonlinear gyrokinetics are plotted for 3 different ITER scenarios.  ITER baseline (Blue), a reduced input power, high Q scenario (Red), and a scenario evaluating a proxy for RMP ELM suppression effects (Green).}
\end{figure}

As demonstrated earlier in this section, the completion of the IBS base case predictions generated a set of surrogate models that have been trained to understand the response of fluxes ($Q_i$, $Q_e$, and $\Gamma_e$) to changes in input variables (such as normalized gradient scale lengths, temperature ratio and collisionality).  Herein lies the power of the surrogate modeling approach.   With trained surrogates in hand, we were able to utilize this model to predict variations around the base case conditions inexpensively.  These surrogate predictions can then be tested with nonlinear gyrokinetic simulations and the results of the high fidelity simulations can be added to the surrogates to improve their training allowing for rapid convergence of profile predictions around the base case conditions.   This ability has been leveraged to look at a potentially higher gain scenario which may be accessible to ITER.  After operating with full input power (53MW) and obtaining the converged profiles shown in Figure \ref{itsum}, it may be possible to lower the total input power, while still maintaining high fusion power, thus increasing the overall plasma gain obtained.  This is possible due to the fact that, as shown in Table \ref{tabsum}, the ITER baseline operates about 45\% above the L to H threshold.  In burning plasma conditions the plasma is, by definition, heated predominately by the fusion alphas.  This state, coupled with the stiff ITG transport that was identified in the previous section, could potentially allow access to a higher Q scenario, simply by decreasing the overall input power to just above the L to H threshold.
\\
\indent
To test this scenario we performed the following gyrokinetic profile predictions.  The input power was scaled from the IBS condition down from 53MW down to 29MW.  This was done simply by decreasing the auxiliary heating powers (NBI and ECH) by the ratio of the total input powers (29/53), while keeping all other aspects of the heating profile unchanged.  Before attempting these simulations, we assessed the potential impact on the pedestal pressure that may result from a drop in the discharges presumed drop in $\beta_N$.  For these conditions, the dependence of the EPED predicted pedestal pressured to changes in $\beta_N$ is extremely weak and therefore for this analysis we assumed that the pedestal pressure would remain unchanged when the input power was dropped.  As is discussed later in this section, we checked the validity of this assumption later after converged profiles were obtained. Although a drop from 53 to 29MW appears to be significant, it is important to note that approximately 100MW of input power is being provided to the plasma via the fusion alphas.  Therefore the drop in overall input power is only approximately 15\% (153 to 129MW).  With an unchanged boundary condition, we performed an additional 3 iterations using the PORTALS framework before obtaining a flux-matched prediction for the new, reduced input power condition.  The profiles obtained from this exercise are plotted in Figure \ref{itall}.  The convergence to a new solution in just 3 additional iterations (15 nonlinear gyrokinetic runs) is phenomenal and demonstrates the power of the surrogate accelerated profile prediction for machine optimization.  

As shown in Figure \ref{itall}, the profiles for the new, low input power scenario (red) are quite similar to those from the base ITER baseline condition (blue).  In fact, only differences in the electron temperature and density are even visible, with the $T_i$ profiles being essentially unchanged within the scale of the plot.  This point emphasizes the nature of the stiff ITG transport in these conditions, as even a 15\% drop in the input power resulted in essentially unchanged ion temperatures.  Because the profiles were relatively unchanged during this variation in input power, the total fusion power generated in this scenario was reduced only slightly from 498MW (IBS) to 489 MW in this new scenario. With a significant reduction in the auxiliary power and only a modest reduction in the overall fusion power, the estimated plasma gain in this condition increased to approximately $Q_{plasma}=17$.  We note that this condition still is predicted to operate about 8$\%$ above the L to H threshold and therefore additional gains may be possible while still maintaining H-mode conditions.   However, there are caveats to this analysis.  1.) the L to H threshold is known to have significant uncertainties and therefore it is not clear how attainable this solution might be in until ITER is in operation, particularly as it concerns to the ability to maintain the H-mode pedestal at the power levels assumed in this study.  2.) We made an assumption of a fixed pedestal during this input power drop based on the observation that the total pedestal pressure was insensitive to changes in $\beta_N$ around the IBS condition.  The total change in the $\beta_N$ from the blue (base IBS profiles) to the low input power profiles (red) in Figure \ref{itall} is extremely small (1.782 to 1.747).  This change was evaluated using the EPED neural network model \cite{meng17} and it was found that the resulting change in the pedestal pressure was estimated to be 0.14$\%$.  We considered this change to be negligible in the context of our work.  However, one could in principle to investigate if these small changes in the pedestal pressure would lead to a significant decrease in Q through an iterative process.  We consider this out of the scope of this paper and it is left for future work. 3.) The outermost radial position that was simulated in this work was r/a = 0.9.  Therefore, any potential profile changes that occur between this location and the top of the pedestal would not be captured in this analysis.
\\
\indent
Since it is well known that the pedestal can play a crucial role in effecting the performance of H-mode conditions, we investigated an additional operational condition to predict the effects on performance.  The predicted ELMs in ITER are known to be large enough to damage the material surfaces of ITER's divertor and first wall.  It is therefore anticipated that ITER will operate with the use of Resonant Magnetic Perturbations (RMPs) that are meant to allow for ELM-free operation while maintaining high performance \cite{eva13}.  While obtaining accurate projections in tokamak plasmas with RMPs is an area of active research, we attempted to evaluate the performance effects that might be obtained from RMP effects on the ITER pedestal.  Motivated by work from Evans and colleagues \cite{eva08}, we assumed that the pedestal density might be degraded down to 75\% of its initial value while leaving the ion and electron temperature pedestals essentially unchanged.  Although crude, this approximation seems to be roughly in line with experimental observations on DIII-D, although we note that more recent work has shown even less of a reduction in pedestal density may be obtained through tailored operation.  With a reduced pedestal density we utilized the existing surrogates and nonlinear gyrokinetic simulation to predict the new performance.  Convergence was obtained with only 6 additional iterations (total of 30 nonlinear gyrokinetic simulations).  The profiles obtained from this exercise are plotted in green in Figure \ref{itall}.  Unsurprisingly the profiles display a more significant variation from the standard ITER baseline condition.  With a lower density, the electron and ion temperature becomes slightly less coupled with $T_e$ increasing more relative to $T_i$.  Again, the striking feature found in Figure \ref{itall} is that the ion temperature profile shape is nearly unchanged even with the significant drop in the density due to the strong ITG transport present.  From a performance standpoint, this condition exhibits slightly higher density peaking, $n_e(0.2)/<n_e> = 1.46$, in line with the drop in effective collisionality at lower density, and therefore consistent with qualitative observations from experimental work \cite{ang09}.  The total fusion power is predicted to be 316MW with 53MW of input power, yielding a $Q_{plasma}$ = 5.98, all while still operating approximately 44$\%$ above the L to H threshold.  Although this is not a rigorous investigation of the effect of RMPs in ITER, it does demonstrate that even with significantly degraded pedestal conditions, like those that may be observed due to RMP application, ITER can still achieve burning plasma conditions. 

\section{Predicting the isotope effect in ITER due to core turbulence}
\begin{figure}[h]
\centering
\includegraphics[width=0.9\linewidth]{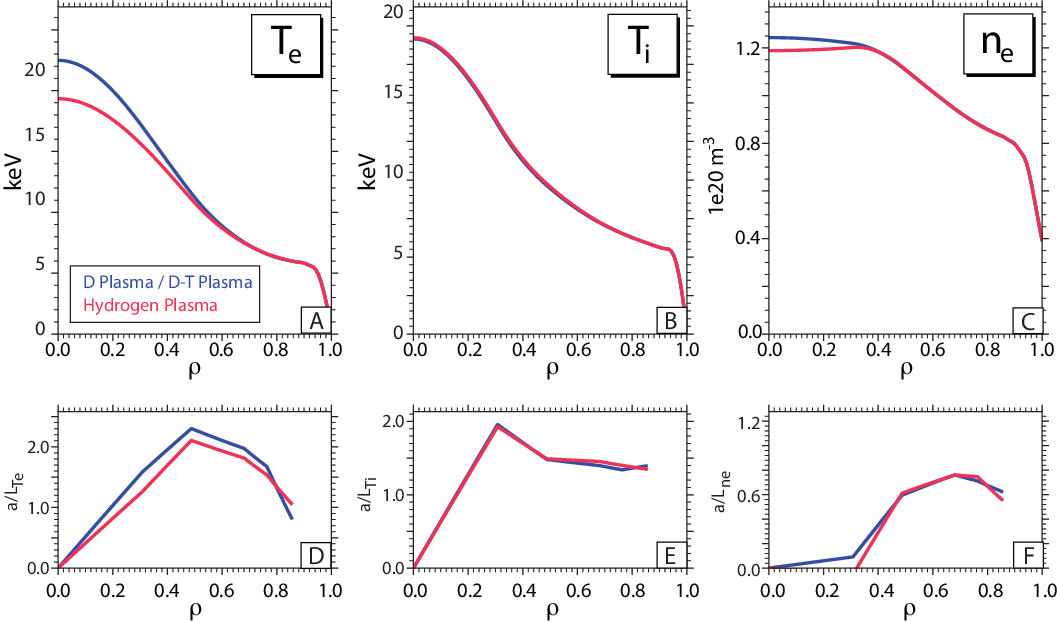}
\caption{\label{iso}(Color Online) Profiles of $T_e$ (A), $T_i$ (B), and $n_e$ (C), as predicted via nonlinear gyrokinetics are plotted for D-T conditions (Blue), D conditions (Blue), and H conditions (Red).  Note that the D and D-T results are the same as there was not statistically meaningful changes in the fluxes when running these cases.  The corresponding normalized gradient scale lengths are plotted in panels D-F.}
\end{figure}

The unique aspect of this work is the ability to predict kinetic profiles of ITER based on nonlinear gyrokinetics.  Prediction of kinetic profiles is relatively uncommon and allows for comparison with a range of 0-D quantities, not typically possible during gyrokinetic studies.  The energy confinement time is one such quantities that has been studied extensively via empirical modeling but only recently have these scalings been compared with high fidelity gyrokinetic predictions.  Tokamaks worldwide have reported observations of a so-called isotope effect, where the energy confinement time is reported to increase with fuel isotope mass, in apparent contradiction to a naive gyro-Bohm scaling which suggests that transport should scale with the fuel ion Larmor radius ($\propto m$).  However, recent theoretical work by Belli et al. suggests that the isotope effect arises from the electron dynamics and is not expected to play a significant role when turbulence is ion dominated \cite{bel20}.  The ITG dominated nature of the turbulence reported in the sections above and the potentially important implications of the isotope effect on ITER performance, motivated additional simulations to probe the isotope effect using nonlinear gyrokinetic profile predictions.   
\\
\indent
The isotope effect in IBS conditions was studied by completing two additional profile predictions where the main ion was changed from a 50/50 mix of deuterium and tritium (m = 2.5 amu) to a pure deuterium plasma (m = 2.0 amu), and a pure hydrogen plasma (m = 1.0 amu).  Because this is a theoretical exercise, aimed only at studying any potential isotope effect arising from the plasma turbulence, the target heat fluxes were kept fixed to the values obtained from the converged D-T condition and the profile for the pure D and pure H cases were then determined through the convergence process described above.  This situation is artificial, since it is unlikely real plasmas could obtain the exact same heating profile, alpha heating would not be present in D and H plasmas at any significant level, and changes in collisional exchange would occur simply from changing the main isotope mass.  However, this theoretical exercise allows us to answer the question of whether or not the turbulence present in the IBS exhibits an isotope effect.  We note that these investigations did not only change the main ion in the simulations, but total number of species was further reduced by grouping all the impurities into a single species (previously W was kept as a separate species with concentration 1.5e-5 $\times n_e$).  This was done to reduce computing resources and the approximation was checked by comparing identical D-T simulations with 5 kinetic species compared to 4 kinetic species (all impurities lumped).  The results from the 4 species run were statistically indistinguishable from the 5 species run indicating this is a fairly robust approximation.  
Note that, because the isotope mass changes, we could not re-use the fluxes surrogates and the simulations from the previous predictions and needed to be completed from scratch. 
The $T_e$, $T_i$, and $n_e$ profiles that were obtained from this exercise are plotted in Figure \ref{iso}.

  As shown in Figure \ref{iso} the profiles obtained for a 50/50 D-T plasma are identical to those obtained from a pure D plasma.  This was determined by the fact that there was no statistically significant change in the fluxes (outside derived uncertainties on the simulation results - typically 10-15$\%$) found when running the D and D-T simulations.  This result in itself suggests that the isotope effect is likely not playing a significant role in the turbulence of the ITER baseline conditions.   In contrast to the pure D plasma results, the hydrogen plasma does exhibit significantly different profiles.  This simulation took approximately 14 iterations to completely converge despite ending up quite close to the D-T profiles.  There is a slight reduction in the core density coming from a hollowing of the density profile near axis (demonstrated in Figure \ref{iso}c and f). Otherwise the primary change occurs in the electron temperature profile, with hydrogen exhibiting lower temperatures than D or D-T plasmas.  Similar to other profile predictions discussed above, the ion temperature profile remained nearly unchanged in this exercise independent of the isotope used.  Again, this is likely a symptom of the very strong ITG transport present in the IBS conditions, which appears to persist independent of the fuel ion studied.   We note that the observed profile changes (drop in the $T_e$ profile, unchanged $T_i$ profile and fairly unchanged $n_e$ profile) are in good agreement with changes observed in isotope experiments on JET \cite{del24} and the results of this exercise clearly disagree with the naive dependence of the energy confinement time on ion mass that is present in the $\tau_{ITER98P(y,2)}$ empirical scaling: $\tau_e \propto m^{0.19}$ \cite{iter99}.

\begin{figure}[h]
\centering
\includegraphics[width=0.75\linewidth]{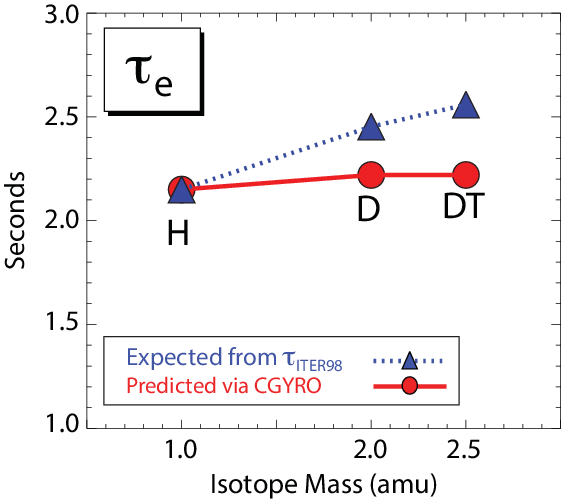}
\caption{\label{tau}(Color Online) The energy confinement times expected from the $\tau_{ITER98P(y,2)}$ empirical scaling: $\tau_e \propto m^{0.19}$ (blue) are compared to those found from the gyrokinetic profile modeling (red).  The value of the hydrogen energy confinement is assumed common between the two and the disagreement is evident by the deviation of the two curves.}
\end{figure}

Assuming that the hydrogen value of the energy confinement is 2.15 seconds (as found from the profiles in Figure \ref{iso}), the empirical scaling would suggest that the D and D-T plasma energy confinement times should be significantly higher, with values in the 2.4 -2.6 second range.  However, from our predictions, these values are identical with a derived energy confinement time of approximately 2.22 seconds.  As shown in Figure \ref{tau}, these results are therefore in clear disagreement with empirical scaling laws and suggest that the turbulence in the ITER baseline should not be expected to exhibit any significant isotope effect.  This result could have significant implications for operation of the device, since such effects have been assumed to exist when projecting ITER performance and operational characteristics.  These results appear inconsistent with the $\tau_{ITER98P(y,2)}$ scaling.  However, recent re-analysis of the ITER H-mode database resulted in potentially only a weak scaling of energy confinement with isotope mass with uncertainties ($\tau_{ITER20} \propto m^{0.2 \pm 0.17}$) \cite{ver21}.  Therefore, the results are potentially in-line with this new analysis.   Furthermore, the results of our work only speak to the role of core turbulence in generating an isotope effect.  Some evidence suggests that the pedestal may be responsible for some of the observed isotope effect \cite{gar22}, and these effects would not be captured with our analysis that assume a fixed boundary condition at $r/a = 0.9$.  Our results are in good agreement with qualitative predictions from the work of Belli and colleagues \cite{bel20} and suggest that ITG dominated plasmas will not observe significant isotope scaling of energy confinement.

\section{Conclusion and Discussion}
This paper presented the first nonlinear gyrokinetic profile predictions of ITER plasmas and utilized the strength of the PORTALS surrogate-accelerated profile prediction techniques to study and optimize ITER performance around the baseline conditions.   Simulations of kinetic profiles for the ITER baseline scenario (IBS) were based on original JINTRAC modeling of the IBS that included core pellet fueling \cite{man20} and included subsequent updates to the profiles that were performed as part of Reference \cite{gri18} and \cite{hol23}, including updated pedestal predictions consistent with EPED and prediction of the rotation profiles via TGLF-SAT2. Nonlinear gyrokinetic simulation utilizing high fidelity, ion-scale physics and 5 gyrokinetic species was used to predict the $T_e$, $T_i$, and $n_e$ profiles that would result from the operation in the IBS.   Notably, these predictions included self consistent alpha heating, radiation losses, and collisional equilibration. Converged profiles were achieved within 14 total iterations ( 5 radial locations x 14 iterations = 70 nonlinear gyrokinetic simulations).  These profiles indicate that the IBS should achieve approximately 500 MW of fusion power with approximately 53MW of total input power, resulting in a plasma gain of Q=9.43.  The predicted energy confinement is in good agreement with the $\tau_{ITER98P(y,2)}$ scaling within the scatter of the database, with a predicted $H_{98} = 0.89$.  It was also found that the predicted density peaking is in relatively good agreement with the data present in the Angioni density peaking database \cite{ang09}.  Transport in these conditions was demonstrated to be dominated by ITG across the profile and the IBS exhibited characteristics of stiff ITG transport. 
\\
\indent
The use of surrogate accelerated, nonlinear gyrokinetic modeling was demonstrated to enable optimization of other ITER scenarios.  The surrogates were able to leverage the training data from the base case conditions to allow for rapid convergence of nonlinear gyrokinetic profile predictions (3-6 additional iterations).  The IBS base conditions were operated approximately $43\%$ above the L to H threshold allowing the total auxiliary power to be reduced to 29MW and still remain above the the L to H power threshold.  The extremely stiff ITG transport was found to essentially pin the $T_i$ profile with only modest changes in the $T_e$ and $n_e$ profiles.  As a result, the plasma gain increased significantly to approximately Q=17 with approximately unchanged fusion power.  This analysis assumed a fixed pedestal condition despite a $\sim 15\%$ decrease in the total heating, which was motivated by the weak dependence of the pedestal pressure on $\beta_N$ around the base conditions.  However, this assumption could be relaxed and investigated in future work.  An additional variation was made around the IBS base conditions.  The pedestal density was dropped to $75\%$ of its nominal value to approximate a potential change in the edge due to RMP application.  Stiff ITG transport continued to play an obvious role with very little change in the $T_i$ profile observed.  However, despite the drop in pedestal density, the scenario still exhibited burning plasma conditions with a Q of approximately 6.0.  These results are generally promising for the prospect of ITER reaching its mission goals and sustaining burning plasma conditions.
\\
\indent
The last investigation covered in this work was related to the isotope scaling of confinement in ITER baseline conditions.  To test whether turbulence in the IBS exhibits any significant isotope effect, the main ion was changed from a 50/50 D-T mix in the gyrokinetic simulations and profiles were predicted for  pure H and D plasmas.  This exercise was performed under the assumption of a completely fixed heating and radiation loss profile with known changes in collisional exchange not included.  The objective of this exercise was to demonstrate whether the turbulence in the IBS would exhibit an isotope effect. It was found that transport in D was statistically indistinguishable from the D-T conditions that were simulated for the IBS.  Changes in the profiles were predicted in H plasmas when compared to D and D-T conditions.  However, these changes were relatively modest and the resulting energy confinement only slightly decreased.  When compared to the $\tau_{ITER98P(y,2)}$ scaling the simulated increase with isotope mass was found to be in clear disagreement with the empirical scaling law.  However, we note that recent re-analysis of the H-mode database suggest that an extremely weak mass dependence of energy confinement is consistent with the database within estimated uncertainties.   However, it is important to emphasize that these results do not indicate that ITER will not exhibit any isotope effect under any operational conditions.  Instead, this analysis is applicable for the turbulence in the IBS specifically and other operational points may see different results.  Furthermore, if the isotope effect primarily arises from the pedestal region, this effect would not be captured by our analysis. Overall the conclusion is that the turbulence in the ITER baseline scenario is unlikely to exhibit a significant core confinement increase (due to changes in transport) as the fuel mix or species is modified.
\\
\indent
The results presented here represent arguably the highest fidelity predictions of the plasma core in ITER baseline conditions ever performed.  These results are generally promising for ITER, which is found to likely meet is mission goals and has potential avenues for reaching higher gain conditions.  Our analysis suggests that increases in energy confinement should not be expected when transitioning from H to D-T plasmas that are similar to the IBS conditions due to the lack of isotope effect exhibited in the dominant ITG turbulence.  Overall, this work provides some insight into ITER's anticipated performance and highlights a promising first-principles based method for predicting core profiles and performance in future devices.

\section{Acknowledgments}
This work was supported by DoE contract numbers, DE-SC0014264, DE-SC0024399, DE-SC0018287, and DE-FG02-95ER54309. The authors would like to thanks Drs. Paola Mantica and Brian Grierson for providing us with access to the original input data used in this work.  We would also like to thank Professor Carlos Paz-Soldan for discussions on RMP effects in current tokamaks and ITER.  This research used resources of the National Energy Research Scientific Computing Center (NERSC), a U.S. Department of Energy Office of Science User Facility located at Lawrence Berkeley National Laboratory, operated under Contract No. DE-AC02-05CH11231 using a NERSC ALCC computing award (m4201).  Some of the simulations performed in this paper were performed on the MIT PSFC Engaging cluster.

\section{References}
\bibliographystyle{unsrt}
\bibliography{bib_new}

\end{document}